\documentclass[12pt]{article}
\usepackage{graphicx}
\usepackage{epsfig}
\usepackage{amsfonts}

\begin{document}

\newcommand{\beq}{\begin{equation}}
\newcommand{\eeq}{\end{equation}}
\newcommand{\beqa}{\begin{eqnarray}}
\newcommand{\eeqa}{\end{eqnarray}}
\newcommand{\beqar}{\begin{eqnarray*}}
\newcommand{\eeqar}{\end{eqnarray*}}
\newcommand{\labell}[1]{\label{#1}\qquad_{#1}} 

\renewcommand{\thepage}{\arabic{page}}   
\setcounter{page}{1}

   
\centerline{\Large \bf Quasilocal Formalism and}   
\vskip 0.4cm   
\centerline{\Large \bf Black Ring Thermodynamics}   
\vskip 1cm   
   
\renewcommand{\thefootnote}{\fnsymbol{footnote}}   
\centerline{
{\bf Dumitru   
Astefanesei${}^{1,2}$
} 
and {\bf Eugen Radu${}^{3}$
}}    
\vskip .5cm   
\centerline{${}^1$\it Harish-Chandra Research Institute,}   
\centerline{\it Chhatnag Road, Jhusi, Allahabad 211019, INDIA}   
\vskip .5cm   
\centerline{${}^2$ \it Perimeter Institute for Theoretical Physics,}   
\centerline{\it Waterlo, Ontario N2L 2Y5, CANADA} 

\vskip .2cm
\centerline { E-mail: dastef@hri.res.in, \,dastefanesei@perimeterinstitute.ca}

\vskip .5cm   
\centerline{${}^3$ \it Department of
Mathematical Physics,}   
\centerline{\it National University of Ireland Maynooth, IRELAND}   

\vskip .2cm
\centerline { E-mail: radu@thphys.nuim.ie }   

\setcounter{footnote}{0}   
\renewcommand{\thefootnote}{\arabic{footnote}}   
   
\begin{abstract}
The thermodynamical properties of a dipole black ring
are derived using the quasilocal formalism.
We find that the dipole charge appears in the first law 
in the same manner as a global charge. Using the Gibbs-Duhem 
relation, we also provide a non-trivial check of the 
entropy/area relationship for the dipole ring. A preliminary 
study of the thermodynamic stability indicates that the 
neutral ring is unstable to angular fluctuations.
\end{abstract}   

\section{Introduction}

Not many objects in physics are as fascinating and intriguing as
black holes. The relationship between thermodynamic entropy and 
the area of an event horizon is one of the most robust and surprising 
results in gravitational physics. Even more surprisingly is the fact 
that $4$-dimensional black holes are highly constrained objects. That 
is, an {\it isolated} electrovac black hole can be characterized, 
{\it uniquely} and {\it completely}, by just three macroscopic 
parameters \cite{nohair}: its mass, angular momentum, and charge.\footnote{The 
classical uniqueness results do not apply to black holes with 
{\it degenerate} horizons.} There are no black objects with an electric 
dipole in four dimensions. The black holes have `smooth' 
horizons (there are no ripples or higher multipoles) and are 
clasically stable. Moreover, for asymptotically flat solutions, 
the event horizons of non-spherical topology are forbidden.

Gravity in higher dimensions --- an important active area in both 
string theory and particle physics --- has a much richer spectrum 
of black objects than in four dimensions. For example, the 
vacuum black ring found by Emparan and Reall in ref.~\cite{Emparan:2001wn} 
has a non-spherical event horizon of topology $S^2 \times S^1$. It was 
also explicitly proved in ref.~\cite{Emparan:2001wn} a `discrete' 
non-uniqueness: There is a range of values for the mass and angular 
momentum for which there exist $three$ solutions, a black hole and 
two black rings. A more dramatic `continuous' violation of `uniqueness' 
was presented in ref.~\cite{Emparan:2004wy}. 
The solution describes a stationary black ring {\it electrically} coupled 
to a $2$-form potential and a dilaton. The ring creates a field analogous 
to a dipole, with no net charge measured at infinity. In this way, a family 
of black rings differing only by their dipole charge is obtained.\footnote{In 
fact this is true at the classical level. Once the charges are quantized in terms 
of the brane numbers, a discrete family of dipole rings is obtained.} Then 
it is clear that, unlike four dimensions, in higher dimensions not all 
black hole equilibrium configurations are completely characterized by a few 
{\it asymptotic} conserved charges. It is not yet known if these solutions are 
stable.

The black ring solutions satisfy the first law of black hole mechanics, 
thus suggesting that their entropy is also one quarter of the event horizon 
area. For dipole black rings, the novelty is that the dipole charge enters 
the first law in the same manner as an ordinary global charge \cite{Emparan:2004wy,Copsey:2005se,Larsen:2005qr}. 
A derivation of the first law of 
black ring solutions based on the Hamiltonian formalism was presented 
in ref.~\cite{Copsey:2005se}.

In this paper, we will take a slightly different route in deriving the first 
law for the dipole ring. Our proposal is to compute the thermodynamical 
quantities by employing the  quasilocal formalism of Brown and York 
\cite{Brown:1992br} supplemented by boundary counterterms. In this way, 
the difficulties associated with the choice of a reference background 
for a rotating spacetime in the presence of matter fields are avoided.

\section{The general framework}
It is well known that the gravitational action
contains divergences even at tree-level --- they arise from integrating over 
the infinite volume of spacetime.  For $5$-dimensional asymptotically flat 
solutions with a boundary topology $S^3\times R$, the action can be regularized 
by the following counterterm \cite{Kraus:1999di}\,: 
\begin{eqnarray}
\label{Ict}
I_{ct}=-\frac{1}{8 \pi G}\int_{\partial M} d^{4} x \sqrt{-h}
\sqrt{\frac{3}{2}\,\mathcal{R}},
\end{eqnarray}
where $\mathcal{R}$ is the Ricci scalar of the induced metric on the
boundary $h_{ij}$.
Varying the total action (which contains the Gibbons-Hawking boundary term)
with respect to the
boundary metric $h_{ij}$, we compute the divergence-free boundary stress-tensor
\begin{equation}
\label{Tik}
T_{ij}=\frac{2}{\sqrt{-h}}\frac{\delta I}{\delta h^{ij}}=
\frac{1}{8\pi G}\Big( K_{ij}-h_{ij}K
-\Psi(\mathcal{R}_{ij}-\mathcal{R}h_{ij})-h_{ij}\Box \Psi+\Psi_{;ij}
\Big),
\end{equation}
where $K_{ij}$ is the extrinsic curvature of the boundary and 
$\Psi=\sqrt{\frac{3}{2\mathcal{R}}}$.
Provided the boundary geometry has an isometry generated by a
Killing vector $\xi ^{i}$, a conserved charge
\beqa
{\frak Q}_{\xi }=\oint_{\Sigma }d^{3}S^{i}~\xi^{j}T_{ij}
\label{charge}
\eeqa
can be associated with a closed surface $\Sigma $ \cite{Mann:2003 Found}. 
Physically, this means that a collection of observers on
the hypersurface whose metric is $h_{ij}$ all observe the same value
of ${\frak Q}_{\xi }$ provided this surface has an isometry
generated by $\xi$. For example, if $%
\xi =\partial /\partial t$ then ${\frak Q}$ is the conserved 
mass/energy $M$.

Upon continuation to imaginary time, the
gravitational thermodynamics is then formulated via the Euclidean
path integral.
The thermodynamic system has a constant temperature $T=1/\beta$ 
which is determined by requiring the Euclidean
section be free of conical singularities.
In a very basic sense, gravitational entropy can be regarded as
arising from the Gibbs-Duhem relation applied to the path-integral
formulation of quantum gravity \cite{Mann:2003 Found}.
The total action $I$ is evaluated from the classical solution to the field
equations, which yields an expression for the entropy
\begin{equation}
S=\beta ({\frak M}-\mu _{i}{\frak C}_{i})-I,  \label{GibbsDuhem}
\end{equation}%
upon application of the Gibbs-Duhem relation to the partition 
function 
\cite%
{Mann:2003 Found} (with chemical potentials ${\frak C}_{i}$ and
conserved charges $\mu _{i}$).
 The first law of thermodynamics is 
then
\begin{equation}
dS=\beta (d{\frak M}-\mu _{i}d{\frak C}_{i}). 
\label{1stlaw}
\end{equation}
However, we will find that
a key point in our intuition about Euclidean sections
does not apply to black rings --- there is no real  {\it non-singular}
Euclidean section in this case.
Nevertheless, as argued in ref.~\cite{Brown:1990fk}, these configurations 
still can be described by a complex geometry  and a real action 
(for other examples, see refs.~\cite{examples1,examples2}).


\section{Asymptotic conserved charges for the dipole black ring}

For a detailed study of the dipole ring we refer the reader to 
ref.~\cite{Emparan:2004wy}, whose notation we follow. The line 
element of this solution is written as
\begin{eqnarray} 
\label{magnetic}
&&ds^2=-\frac{F(y)}{F(x)}\left(\frac{H(x)}{H(y)}\right)^{N/3}
\left(dt+C(\nu,\lambda)\: R\:\frac{1+y}{F(y)}\: d\psi\right)^2\\[3mm]
&&+\frac{R^2}{(x-y)^2}\: F(x)\left(H(x)H(y)^2\right)^{N/3}\left[
-\frac{G(y)}{F(y)H(y)^N}d\psi^2-\frac{dy^2}{G(y)}
+\frac{dx^2}{G(x)}+\frac{G(x)}{F(x)H(x)^N}d\varphi^2\right]\,,\nonumber
\end{eqnarray}
where
\begin{eqnarray}
F(\xi)=1+\lambda\xi,~~~ G(\xi)=(1-\xi^2)(1+\nu\xi),~~~
H(\xi)=1-\mu\xi\, ,
\end{eqnarray}
and $C(\nu,\lambda)=\sqrt{\lambda(\lambda-
\nu)(1+\lambda)/(1-\lambda)}$.
The 
constant $N$ which enters the above relations is related to the 
dilaton coupling constant $\alpha$ through
$N=(\alpha^2/4+1/3)^{-1}$. The values $N=1,2,3$ are of particular 
relevance to string and M-theory. 

The coordinates $x$ and $y$ vary within the ranges
$-1\leq x\leq 1,- \infty<y\leq -1, 
$
while $R,\lambda,\mu$ and $\nu$ are real parameters with 
 $0< \nu\leq\lambda<1$.
To avoid conical singularities at $x=-1$ and $y=-1$ one sets
\begin{eqnarray}
\label{periodN} 
\Delta\psi=\Delta\varphi=
2\pi\frac{(1+\mu)^{N/2}\sqrt{1-\lambda}}{1-\nu}, 
\end{eqnarray} 
while the
singularity at $x=+1$ is avoided by requiring 
\begin{eqnarray}
\label{equilN}
\frac{1-\lambda}{1+\lambda}\left(\frac{1+\mu}{1-\mu}\right)^N=
\left(\frac{1-\nu}{1+\nu}\right)^2\,. 
\end{eqnarray} 
With these choices, the solution has a regular horizon
at $y=-1/\nu$,
of topology
$S^1\times S^2$ and area
$
{{\cal A}_H}=8\pi^2 R^3
(1+\mu)^N\nu^{(3-N)/2}(\mu+\nu)^{N/2}
\sqrt{\lambda(1-\lambda^2)}/(1-\nu)^2(1+\nu)~,
$
 an ergosurface of the same topology
at $y=-1/\lambda$, and an inner spacelike singularity at
$y=-\infty$.
 Asymptotic spatial infinity is reached as $x\to y\to
-1$.

The dilaton $\tilde \phi$ and the two-form potential are given by
\begin{eqnarray}
\nonumber
e^{\tilde \phi}=\left(\frac{H(x)}{H(y)}\right)^{N\alpha/2}, ~~~~
B_{t\psi}=C(\nu,-\mu)\:\sqrt{N}\:R\:(1+y)/H(y)+k,
\end{eqnarray}
where $k$ is a constant. The main observation in 
ref.~\cite{Copsey:2005se} is that the constant $k$ is 
not arbitrary. 
Usually, the gauge potential  is
globally defined and non-singular everywhere outside 
(and on) the horizon. 
However, Copsey and Horowitz have 
shown that this is incompatibile with the assumptions 
that the dipole charge is nonzero and that $B$ is invariant 
under the spacetime symmetries. The constant $k$ must be chosen so 
that $B_{t\psi}(y=-1)=0$ and implies that $B_{\mu\nu}$ necessarily 
diverge at the horizon. For our analysis it is important 
to note that this is just a purely gauge effect --- the 
physical field $3$-form $H=dB$ remains finite at the horizon. 

To evaluate asymptotic expressions at spacelike infinity, it is
convenient to introduce coordinates in which the asymptotic flatness
of the solution becomes manifest.
 Our choice for this transformation is
\begin{eqnarray}
\label{transf} 
x=1-\frac{2r^2}{r^2+R^2\cos^2 \theta}, ~~
y=1-\frac{2(r^2+R^2)}{r^2+R^2 \cos^2 \theta},
\end{eqnarray}
$r$  corresponding to a normal coordinate on the boundary, $0\leq
r<\infty$, $0\leq \theta\leq \pi/2$. In these coordinates, the black 
ring approaches asymptotically the Minkowski background
$ds^2= d \bar{r}^2+\bar{r}^2 \Big(d \theta^2
+\sin^2\theta d \bar{\psi}^2+ \cos^2\theta
d\bar{\varphi}^2\Big)-dt^2,$
where $\bar{\varphi}$ and $\bar{\psi}$ are angular coordinates
rescaled according to eq.~(\ref{periodN}) and 
$\bar{r}=((1-\lambda)/(1-\nu))^{1/2}r$.  

The mass $M$ and angular momentum $J$ of the black ring solution 
can be computed by employing the quasilocal formalism.
The relevant components of the boundary stress tensor are
\begin{eqnarray}
\nonumber
&&8 \pi G T_\psi^t=\frac{4R^3C(\nu,\lambda)}{(1-\lambda)}
\sqrt{\frac{(1+\mu)^{-N}(1-\nu)}{1-\lambda}}\frac{\sin^2 \theta}{r^3}+O(1/r^4),
\\
\nonumber
&&8 \pi GT_t^t=\frac{R^2(1+\mu)^{-1-N/2}}{(1-\lambda) }
\Big(N\mu+\lambda(3-(N-3)\mu)\Big)\sqrt{\frac{1-\nu}{1-\lambda}}\frac{1}{r^3}+O(1/r^4).
\end{eqnarray}
Using eq.~(\ref{charge}) we obtain the following expressions for 
mass and angular momentum of the dipole ring solution
\begin{eqnarray}
\label{mass} 
M&=&\frac{3\pi R^2}{4G
}\frac{(1+\mu)^N}{1-
\nu}\left(\lambda+\frac{N}{3}\frac{\mu(1-\lambda)}{1+\mu}\right)\,,
\\
\label{J}
J&=&\frac{\pi R^3}{2G
}\frac{(1+\mu)^{3N/2}\sqrt{\lambda(\lambda-
\nu)(1+\lambda)}}{(1-\nu)^2},
\end{eqnarray} 
which match the ADM values computed in ref.~\cite{Emparan:2004wy}.
\section{The dipole ring action}
The partition function for the gravitational field is defined by a sum
over all smooth Euclidean geometries which are periodic with period
$\beta$ in imaginary time.
This integral is computed by using the saddle-point approximation.
The energy and entropy are evaluated by standard thermodynamic relations.
Naively, one may expect to find a real Euclidean section
for a black ring solution by using the
analytical continuation 
$t \to i\tau$ supplemented with $C\to i\bar{C}$.
However, it can  be verified 
that the conical singularities at  $x=\pm 1$, $y=-1$ of the
Euclidean line element cannot be removed for any choice of 
$(\lambda,\nu,\mu)$ which assures a real $\bar{C}$.
Therefore, we are forced to work with a complex geometry.

We adopt here the `quasi-Euclidean' method of 
ref.~\cite{Brown:1990fk} in which the Wick transformations 
affect the intensive variables, such as the lapse and shift 
($N\rightarrow -iN$ and $N^k\rightarrow -iN^k$), but for which
the extensive variables (such as energy) remain invariant. It is 
important to be mentioned that the Cauchy data and the 
equations of motion remain invariant under this 
complexification.\footnote{Note that since the energy and angular 
momentum are described by three-surface integrals over the Cauchy data, 
they remain invariant and real under this complexification.}
Starting with the ring metric in the canonical (ADM) form we 
obtain the following `quasi-Euclidean' section:
\begin{eqnarray} 
\nonumber
ds^2=N^2\,d\tau^2 +
\gamma_{ij}\,(dy^i-i\,N^i\,d\tau)(dy^j-i\,N^j\,d\tau) \, .
\end{eqnarray} 
No singularities are found on this `quasi-Euclidean' section --- the 
conical singularities at $y=-1,~x=\pm 1$ are avoided 
by taking the same periodicity for $\psi$ and $\varphi$ together with 
eq.~(\ref{equilN}). One has also to identify $\tau$ with a period $\beta$ 
to make the metric regular on the horizon. 
A detailed analysis shows 
that the periodicity  $\beta$  and the shift vector at the horizon 
 $N^{\psi}_H$  reproduce the inverse of the Hawking temperature and 
the angular velocity of the horizon, respectively, as computed 
in ref.~\cite{Emparan:2004wy}
\begin{eqnarray}
\beta=
\frac{4\pi R(\mu+\nu)^{N/2}}{\nu^{(N-1)/2}(1+\nu)}
\sqrt{\frac{\lambda(1+\lambda)}{1-\lambda}},~~~
\Omega_H=\frac{1}{R}\frac{1}{(1+\mu)^{N/2}}\sqrt{\frac{\lambda-
\nu}{\lambda(1+\lambda)}}.
\end{eqnarray}
It is convenient to use the $r,~\theta$ coordinates, as defined by 
eq.~(\ref{transf}), 
in order to compute the boundary terms (Gibbons-Hawking plus the counterterm) 
contribution to the total action.
In the large $r$ limit we find the following finite expression:
\begin{eqnarray}
\label{Ibound} I_{\partial B}= \frac{\pi^2 R^3}{3G(1-\nu^2)}
\Big(3\lambda \sqrt{\frac{\lambda \nu (1+\lambda)}{(1-\lambda)}}
+\sqrt{\frac{\lambda (1+\lambda)}{(1-\lambda)}} (1+\mu)^{N-1}
 (-N\mu
\\
\nonumber
+\lambda (-3+N)\mu))\nu^{1/2-N/2} (\mu
+\nu)^{N/2}\Big).
\end{eqnarray}
The tree-level bulk action is computed by using the trace of the 
Einstein equations 
\begin{eqnarray}
\label{actionH1} 
I_B= \frac{1}{16\pi G}\int_M d^5x\sqrt{-g}{\left(
R-\frac{1}{2}(\partial\tilde\phi)^2-\frac{1}{12}e^{-\alpha\tilde\phi}
H^2\right)}=
-\frac{1}{16\pi G}\int
d^5x\sqrt{-g} e^{-\alpha\tilde\phi}\frac{1}{9}H^2.
\end{eqnarray}
This volume integral evaluated on the Euclidean section takes a simple 
form expressed in terms of the dipole 
charge $q$ and the potential $\Phi$ defined as  
\begin{eqnarray}
\label{elcharge} 
&&q=\frac{1}{4\pi}\int_{S^2}e^{-\alpha\tilde\phi}\ast H=
R\sqrt{N}\frac{(1+\mu)^{(N-
1)/2}\sqrt{\mu(\mu+\nu)(1-\lambda)}}{(1-\nu)\sqrt{1-\mu}},
\\
\label{phiN}
&&\Phi=\frac{\pi}{2G
}\left(-B_{t\bar\psi}(y=-1/\nu)\right)
 =\frac{\pi R}{2G }\sqrt{N}\frac{(1+\mu)^{(N-1)/2}\sqrt{\mu(1-\mu)(1-
\lambda)}}{\sqrt{\mu+\nu}}, 
\end{eqnarray}
where $S^2$ is a surface of constant $t,y$ and $\psi$ in the metric
(\ref{magnetic}). Then, the bulk contribution is 
$I_B=\frac{2}{3}\beta q \Phi.$
It is important to precisely point out the nature of the dipole 
charge. A string naturally couples to a $2$-form gauge potential. 
The special case $N=1$ is the NS sector of low-energy string theory.
Then, a fundamental string that carries {\it electric} Kalb-Ramond 
charge is a solution of our theory. The string charge is 
localized on the string and the charge density can be 
visualized as a current on the string. Since the string winds around 
a {\it contractible} circle, no monopole term will appear in the multipole 
expansion for the field. Therefore, the {\it local} charge\footnote{It 
is well defined due to the field equation $d(e^{-\alpha\tilde\phi}\ast H)=0$ (see, 
$e.g.$, ref.~\cite{don} for a nice discussion on different notions of charge).}
(\ref{elcharge}) has a natural interpretation as a source of the dipole 
field. 
 
The total action $I$ is given by
\begin{eqnarray}
\label{Itot}
I=I_B+I_{\partial B}=\frac{\pi^2 R^3}{4G(1-\nu)}(1+\mu)^{N-1}\big(\lambda(1+\mu)-N\mu(1-\lambda)\big),
\end{eqnarray}
which is a strictly positive quantity. For a grand-canonical ensemble ($i.e.$ for 
fixed temperature, angular velocity, and gauge potential), using 
the definition of the Gibbs potential $G(T,\Omega_H ,\Phi )=I/\beta$, 
eq.~(\ref{J}) for the angular velocity, and eq.~(\ref{phiN}) for the 
potential $\Phi$, we obtain
\begin{equation}
\label{Gib} G(T,\Omega_H ,\Phi )=M-\Omega_H J-T S-\Phi q,
\end{equation}
which means that $G(T,\Omega_H ,\Phi )$ is indeed the Legendre
transformation of the energy $M(S,J,q)$ with respect to
$S$, $J$, and $q$. 
The entropy $S=-\left(  \partial G/\partial
T\right) _{\Omega_H \Phi }$ is one quarter of the event horizon area 
$\cal{ A}_H$.
A straightforward calculations  shows that the extensive thermodynamical
quantities
\begin{equation}
J=-\left( \frac{\partial G}{\partial \Omega_H }\right) _{T\Phi },
~~~ q=-\left( \frac{\partial G}{\partial \Phi }%
\right) _{T\Omega_H },
\end{equation}
turn out to coincide with the expressions (\ref{J}) and
(\ref{elcharge}), the first law of thermodynamics
(\ref{1stlaw}) also being satisfied.
\\
\section{Discussion}
Black rings provide a novel theoretical laboratory for studying the 
physics associated with event horizons. Using a counterterm-like method 
for flat spacetimes, we have explicitly shown that the first law of black 
dipole ring mechanics expresses the conservation of energy by relating 
the change in the dipole ring mass $M$ to the change in its area $\cal{ A}_H$, 
angular momentum $J$, and the dipole $q$. An 
extended form of the zeroth law implies that not only the surface gravity, 
but also the other intensive quantities (in our case, the angular velocity and 
the conjugate potential of the dipole charge) should be constant over the 
event horizon. Indeed, we found that the potential $\Phi$ in eq.~(\ref{phiN}), 
which is constant over the event horizon, appears as the conjugate potential 
of the dipole charge in the first law.

The dipole ring does not have a real non-singular Euclidean section. To remedy 
the situation, following ref.~\cite{Brown:1990fk}, we constructed a complex metric 
that transformed the lapse function and the shift vector to imaginary quantities, but 
which kept the gravitational Cauchy data invariant (and hence the extensive quantities). 
The horizon is described by the `bolt' in this complexified geometry but has invariant features 
($e.g$, area, gravity surface, etc.) as in the Lorentzian sector. Then, 
the key features of the physical Lorentzian dipole ring have been preserved.

The analysis of the thermodynamic stability of the black ring solutions
turns out to be very complicated,
simple results being possible in the vacuum case only. 
For $\mu=0$, 
the Euclidean regularity at the horizon, which is equivalent to the
condition that the black ring is in thermodynamical equilibrium,   
gives the equation of state 
\begin{equation}
T=\frac{1}{16 \pi}\sqrt{\frac{\pi^2}{J^2\Omega_H^2}-16\Omega_H^2},
\end{equation}
and the Gibbs potential can be written as
\begin{equation}
G[T,\Omega_H]=\frac{1}{64 \pi G~T^2}\left(-1+\sqrt{\frac{16\pi^2 T^2}{\Omega_H^2}+1}\right).
\end{equation}
The analysis of refs.~\cite{Emparan:2001wn, Emparan:2004wy} reveals
the existence of two branches of solutions
in terms of the dimensionless reduced spin $j$ and  reduced area of the
horizon $a_H$, with
$j^2\equiv \frac{27\pi}{32 G}\frac{J^2}{M^3}=(1+\nu)^3/8\nu,~~
  a_H\equiv \frac{3\sqrt{3}}{16\sqrt{\pi}}\frac{{\cal A}_H}{(GM)^{3/2}}
  =2\sqrt{\nu(1-\nu)}, 
$
which join for $(j^2,a_H)=(27/32,1)$ (corresponding to $\nu=1/2$),
and which are dubbed `large' and `small' according to their area.
For a grand canonical ensemble, the control parameter is $4 \pi T/\Omega_H$, the
value for which the two branches join being $\Omega_H=4\pi T/\sqrt{3}$. 
 
To discuss the thermodynamic stability in a grand canonical 
ensemble, we consider first the specific heat at constant 
angular velocity at the horizon 
\begin{equation}
C_\Omega=T\left(\frac{\partial S}{\partial T}\right)_{\Omega_H}.
\end{equation}
It turns out that only the `large'
black ring solutions with $\Omega_H<4\pi\sqrt{2/3^{\frac{1}{2}}}T$ 
(corresponding to $\nu<2/\sqrt{4+\sqrt{3}}$)  are stable 
against thermal fluctuations,
$C_\Omega>0$. 
When considering instead a canonical ensemble with 
$F[T,~J]=M-TS$, one finds that the specific heat at 
constant angular momentum is always positive
\begin{equation}
C_J=T\left(\frac{\partial S}{\partial T}\right)_{J}> 0,
\end{equation}
which implies the ensemble is thermally stable.
Another `response function' of interest is the 
`isothermal permittivity' $\epsilon_T \equiv \left(\partial J/ 
\partial\Omega_H\right)_{T}$. Since it is always negative, 
the neutral black string solution is unstable to angular 
fluctuations, both in a macrocanonical and canonical ensemble.

At this end, we would like to comment on our counterterm prescription
for the asymptotically flat spacetimes. Unlike asymptotically anti-de 
Sitter (AdS) spaces, the {\it locality} of the counterterm is not 
a priori mandatory for the asymptotically flat spaces 
\cite{Kraus:1999di} --- though, for our purpose it was sufficient 
to consider a local counterterm. We have only investigated stationary 
spacetimes and so, for each value of the cut-off, the slice with the 
induced metric $h_{ij}$ is stationary. Since $\xi=\partial /\partial t$ 
is a Kiling vector of the cut-off boundary, the energy is conserved. 
However, it would be interesting to find  a more general result for 
any asymptotically flat spacetime. It is also worth exploring the 
connection between the holographic charges and the various alternative 
definitions of conserved charges in asymptotically flat spacetimes 
(for AdS, see refs.~\cite{holo1,holo2,holo3}). These issues are 
currently being investigated \cite{donrobb}.

A detailed analysis of the thermodynamics of black ring solutions 
will be presented in a companion paper \cite{noi}.
\\

\textbf{Acknowledgements}\\
We are grateful to Roberto Emparan for fruitful discussions and 
collaboration in the initial stage of this project. We would like 
to thank Robert Mann and Don Marolf for shearing with 
us some of their related ideas on quasilocal formalism, and for 
valuable conversations. We would also like to thank Vijay Balasubramanian, 
Oscar Dias, Henriette Elvang, Greg Jones, David Mateos, Rob Myers, Ashoke 
Sen, and Cristi Stelea for enjoyable discussions during the course of this 
work, and Greg Jones for proof-reading an earlier draft of this paper.

DA thanks the organizers of the Fields Institute Workshop on Gravitational 
Aspects of String Theory, the Perimeter Institute School on Strings, 
Gravity and Cosmology, and String 2005 for stimulating environments.
DA was supported by the Department of Atomic Physics, Government of
India and the visitor programme of Perimeter Institute.
The work of ER was supported by Enterprise--Ireland Basic
Science Research Project SC/2003/390 of Enterprise-Ireland.


\end{document}